\documentclass[12pt]{iopart}
\usepackage{epsfig}
\usepackage{citesort}

\newcommand{\gtrsim}{\,\rlap{\lower3.7pt\hbox{$\mathchar\sim$}}
\raise1pt\hbox{$>$}\,}
\newcommand{\lesssim}{\,\rlap{\lower3.7pt\hbox{$\mathchar\sim$}}
\raise1pt\hbox{$<$}\,}

\newcommand{\beq}{\begin{equation}}
\newcommand{\eeq}{\end{equation}}
\newcommand{\bea}{\begin{eqnarray}}
\newcommand{\eea}{\end{eqnarray}}
\newcommand{\lpart}{\raise.3ex\hbox{$\stackrel{\leftarrow}{\partial}$}}
\newcommand{\rpart}{\raise.3ex\hbox{$\stackrel{\rightarrow}{\partial}$}}
\newcommand{\ldr}{\raise.3ex\hbox{$\stackrel{\leftarrow}{\delta^r}$}}

\begin{document}

\title{Searching for a holographic connection between dark energy and the low-$l$ CMB multipoles}

\author{Kari Enqvist\footnote{kari.enqvist@helsinki.fi}}
\address{University of Helsinki, Department of Physical Sciences \\
and \\
Helsinki Institute of Physics, P.O. Box 64, FIN-00014 University of Helsinki, Finland}
\author{Steen Hannestad \footnote{hannestad@fysik.sdu.dk}}
\address{Physics Department, University of Southern Denmark
Campusvej 55, DK-5230 Odense M, Denmark}
\author{Martin S. Sloth\footnote{sloth@physics.ucdavis.edu}}
\address{Department of Physics, University of California, Davis, CA 95616, USA}

\date{9 Sept. 2004}

\begin{abstract}
We consider the angular power spectrum in a finite universe
with different boundary conditions and perform a fit to the CMB, LSS
and supernova data. A finite universe could be the consequence of a
holographic constraint, giving rise to an effective IR cutoff at
the future event horizon.  In such a model there is a cosmic duality
relating the dark energy equation of state and the power spectrum,
which shows a suppression and oscillatory behaviour that is found
to describe the low $l$ features extremely well. However, much of the
discussion here will also apply if we actually live inside an
expanding bubble that describes our universe. The best fit to the CMB
and LSS data turns out to be better than in the standard $\Lambda$CDM
model, but when combined with the supernova data, the holographic
model becomes disfavored. We speculate on the possible implications.
\end{abstract}

\maketitle

\section{Introduction}

The spatial geometry of the universe is known to be flat with $\Omega=1.02\pm 0.02$
\cite{Bennett:2003bz,Spergel:2003cb} so that even if the geometry is closed, one would
argue that its radius should be much larger than the present
Hubble scale. However, it is still possible that the universe is
relatively small and finite. The simplest example is provided by a
geometry which is not simply connected, such as a torus. In such a
case the universe is effectively a box with periodic boundary
conditions, resulting in suppression of power at large angular
distances as well as in multiple images of the same sources in
different directions. The present CMB data constrains the
fundamental domain to be larger than 1.2 to 1.7 times the distance
to the decoupling surface, depending on model assumptions
\cite{topolimits}.

A more subtle and potentially more far-reaching possibility is
related to holography \cite{holo}. The holographic principle
emerged first in the context of black holes, where it was noted
that a local quantum field theory can not fully describe black
holes while preserving unitarity; for a sufficiently large volume
the entropy of an effective field theory will violate the
Bekenstein bound \cite{bekenstein}. This indicates that a local
field theory overcounts the true dynamical degrees of freedom of a
gravitating system. Therefore, an effective field theory which
correctly accounts for the coarse grained interactions of the true
dynamical degrees (e.g. strings), should be subject to global
constraints that remove the overcounting. This should hold true
also for a field theoretical description of the universe as a
whole.

The issue then is, how would the holographic constraints manifest
themselves in cosmology? At present no rigorous answer exists, but
there are several phenomenological ideas \cite{Banks:2001px}. For
instance, it has been
suggested that the effective field theory should exclude those degrees
of freedom that will never be observed, giving rise to an IR cutoff
$\tilde L$ at the future event horizon
\cite{Banks:2000fe,Li:2004rb,Huang:2004ai,Huang:2004wt,kems}. In a
universe dominated by dark energy, the size of the future event
horizon will tend to a constant which, given the WMAP results on dark
energy, is of the order of $H^{-1}_0$, the present Hubble radius.
Therefore, the consequences of such a cutoff could well be visible at
the largest observable scales and in particular in the low CMB
multipoles, where instead of continuous wave numbers one should deal
with discrete ones even though, strictly speaking, the universe need
not be finite.

In the case of topologically finite universes, the size of the
universe correspond to a constant comoving size, which would be
less that the Hubble radius during the inflationary epoch. This
problem is also avoided if the effective cutoff is associated with
the future event horizon which corresponds to an almost constant
physical size and is never less than the Hubble radius.

The appearance of an IR cutoff may not be the only consequence of
holography. An effective field theory that can saturate the
Bekenstein bound will in general include states with a
Schwarzschild radius much larger than the size of the cutoff
volume. It seems reasonable that one should require
\cite{Cohen:1998zx,Thomas:pq} that the energy in a given volume
should not to exceed the energy of a black hole, which results in
a constraint on the size of the zero point fluctuation, i.e. an
effective UV cutoff. In that case the observations should reveal a
correlation between dark energy and the power spectrum
\cite{kems}.

The universe constrained by holography is finite only in an
effective, field theoretical sense. Therefore it is not obvious
what sort of boundary conditions one should impose. In fact,
periodic boundary conditions would appear to be the least natural.
Instead, one could consider Dirichlet boundary conditions with
quantum fields vanishing at the cutoff scale, or Neumann boundary
conditions with vanishing derivatives ensuring that no currents
flow beyond the cutoff. This bears some resembles to the ``brick
wall'' model \cite{'tHooft:1984re} or the ``stretched horizon''
model \cite{Susskind:1993if} of black holes, where one imposes a
boundary condition for the fields on a plane close to the horizon.
Also in the case where we actually live inside an expanding
bubble, these choices of boundary conditions appear more natural.
The choice of the boundary condition is important because it
determines the way the wave numbers are discretized, which in turn
makes a difference when fitting the actual data. Note also that
many of the signatures of a finite universe considered in the
literature, such as circles-in-the-sky, are specific to periodic
boundary conditions.

No matter what the boundary conditions, the observable consequences of
a holographic constraint would be likely to show up at the largest
observable scales. Intriguingly enough, there are a number of features
in the large angle CMB power spectrum, the most notorious being the
suppression of the quadrupole and the octupole \cite{Bennett:2003bz,Spergel:2003cb}.  In
addition, there are some hints of an oscillatory behaviour of the
power spectrum at low $l$ \cite{Tocchini-Valentini:2004ht}. There has
been attempts to
explain these glitches by trans-Planckian physics
\cite{transplanck,Hannestad:2004ts} and by models of multiple
inflation \cite{huntsarkar}. While the
suppressed quadrupole and octupoles can be explained by such models
with modified primordial power spectra, the temperature-polarization
power spectrum measured by WMAP is {\it higher} than expected and in
general the fit to data cannot be significantly improved by modifying
the primordial power spectrum \cite{Hannestad:2003zs}. The features
seen at higher $l$ are even more difficult to fit by modifying the
underlying power spectrum because they are very localized in
$l$-space.  It is therefore an interesting question whether the low
$l$ features could actually tell us something about holography.

The purpose of the present paper is to consider a discrete power
spectrum with different boundary conditions and make a fit to the
CMB, the large scale structure (LSS) and supernova (SN) data. We
pay a particular attention to the features in the power spectrum
at low $l$. These considerations do not as such require holography
but simply test the possible varieties of discreteness of the data
as dictated by different boundary conditions. To relate this to the holographic
ideas, we focus on a simple toy model \cite{kems} that links the
equation of state of the dark energy to the features at low
multipoles in the CMB power spectrum through an IR/UV cosmic
duality.

The paper is organized as follows. In Sect. II we discuss the various
boundary conditions and the procedure of discretization of the power
spectrum. We deal primarily with Dirichlet and Neumann boundary
conditions. In Sect. III we present the results of fits to the data,
keeping the dark equation of state and the IR cutoff independent. In
this way the results apply more generally, also to the case where we
actually live inside a slowly expanding bubble. As an application to
holography, in Sect. IV we consider a toy model of cosmic IR/UV
duality \cite{kems} which predicts the equation of state for
dark energy as a function of IR cutoff
\cite{Hsu:2004ri,Li:2004rb}. Sect. V contains a discussion
of the results.

\section{Boundary Conditions}

Let us start by assuming that the field theory contains an IR cutoff
$\tilde L$, which we wish to translate into a cutoff at physical
wavelengths.  We will assume an isotropic universe so that the system
is spherically symmetric. As a consequence of the IR cutoff the
momenta $k$ will be discrete, but the allowed wave numbers will depend
on the boundary conditions. However, if the discretization is due to a
holographic constraint, there is no obvious way to choose one
particular boundary condition over another.  In the holographic case
the cutoff $\tilde L$ is only effective: we do not live in any actual
physical cavity and space can well be infinite, even if the effects
are very similar to the effects of living inside a bubble.  Rather,
the nature of the boundary condition would be a manifestation of the
holographic constraint itself and hence at the present
phenomenological level undetermined. Therefore it makes sense to
consider several likely possibilities.

Let
us therefore consider quantization in a spherical potential well with
infinitely high potential walls. The radial
solutions for the wave functions are spherical Bessel functions with
the ground state $j_0\propto \sin(kr)/kr$ with $r<r_B$, where
$r_B$ is the radius of the spherical potential and is related to the IR cutoff.
Imposing different boundary conditions on these solutions result in differences
than can have observable consequences.

Periodic boundary conditions have been extensively discussed in the
literature in the context of non-simply connected spaces. In general
they lead to geometric patterns in the sky that are highly constrained
by data \cite{topolimits}. For Cauchy boundary conditions, we would
need to make
model dependent assumptions about physics beyond the IR cutoff
$\tilde L$, such as that the fields are
exponentially suppressed for $r>\tilde L$. Therefore we prefer to focus on
the unambiguous cases of the Dirichlet and Neumann boundary conditions.

\subsection{Dirichlet}

Requiring that the
solution vanishes at $r=r_B$ one finds that the wavelength of the
ground state is $\lambda_c = 2 r_B$. Thus the IR cutoff $\tilde L$
corresponds to a physical cutoff given by $\lambda_c =
2\tilde L$.

For $x\gg 1$ we may write the Bessel function as
$j_l(x)=\frac{1}{x}\sin\left(x-\frac{l\pi}{2}\right)$.
Thus the allowed wavenumbers, $k_{nl}$, are determined by
 \beq
j_l(k_{nl}r_B)=\frac{1}{k_{nl}r_B}\sin\left(k_{nl}r_B-\frac{l\pi}{2}\right)=0~,
 \eeq
which implies
$
k_{nl}r_B=\frac{l\pi}{2}+2\pi n$ for $n,l\in \mathcal{N}
$,
or
 \beq \label{knl}
k_{nl}=\frac{1}{r_B}\left(\frac{l\pi}{2}+\pi n\right)~.
 \eeq
We emphasize that for each choice of the angular variable $l$, there is a
 different discrete set of the allowed momenta $k$.

\subsection{Neumann}

In this case we require the derivative of the allowed solutions
vanish on the wall of the effective cavity defined by the IR cutoff so
that there
is no current flow out of the observable universe. The radial
dependent part of the solutions is given by the spherical Bessel
functions, so it amounts to requiring that the derivative of the
spherical Bessel function vanishes on the wall of the bubble. Using the
recurrence formulae for Bessel functions, one finds
\beq
\frac{d}{dx}j_l(x)=lx^{-1}j_l(x)-j_{l+1}(x)
\eeq
so that the allowed momenta are defined by the condition
\beq
l(kr_B)^{-1}j_l(kr_B)-j_{l+1}(kr_B)=0~.
\eeq
for all $l,n \in \mathcal{N}$.

\section{Discretization of the power spectrum}

\subsection{Preliminaries}

In our case the discretization of the power spectrum differs
from the more usual case of toroidal universes. Let us therefore repeat some of
the basic steps involved in the discretization procedure.

We need to determine what is the proper relation between the
discrete power spectrum and the usual continuous one. In a finite volume
we can always
expand a wave function $\psi(x)$ in terms of a complete set of orthonormal
solutions to the equation of motion $u_k(x)$ so that
$
\psi(x) = \sum_n \psi_n u_{k_n}(x)
$
with
$
\int |u_{k_n}(x)|^2dx = 1~
$
and
$
\psi_n=\int \psi(x)u_{k_n}^*(x) dx~.
$
The probability of finding the state $\psi$ with momentum $k_n$ is then given
by
\beq
P(k_n)=|\psi_n|^2~.
\eeq
The continuous limit is taken by letting the volume to go to
infinity. The
probability of finding the state $\psi$ with momentum in the interval
between
$k$ and $k+dk$ is then given by
$
P(k)dk= |\psi(k)|^2 dk
$
and
\beq
\psi(k)=\psi_k\sqrt{\frac{dN}{dk}}~
\eeq
where $dN/dk$ is the density of states.
By renormalizing $u_{k_n}(x)$ we may write
$
\psi(k)=\int \psi(x)u^*(k,x)dk
$
where
\beq
u(k,x)=u_{k}(x)\sqrt{\frac{dN}{dk}}~.
\eeq
Hence the continuous limit is reached by the replacement
\beq
\sum_n \to \int dN =\int dk \frac{dN}{dk}~.
\eeq

In a spherically symmetric cavity,
 it is convenient to write the orthonormal solutions in the form
  \beq
u_{nlm}(\vec{x})=A_nN_lj_l(kr)Y_{lm}(\theta,\phi)~,
  \eeq
where $A_n^2$ is the density of states, $N_l$ is an $l$-dependent
normalization factor;
  $r=|\vec{x}|$ and $\Omega_x=(\theta,\phi)$ is the angular direction
  of the coordinate $\vec{x}$. The normalization is chosen such that
 \beq
\int_0^Ldr r^2 \left[A_nN_l j_l(k_nr)\right]\left[A_mN_l j_l(k_mr)\right] =\delta_{nm}~.
 \eeq
According to the previous subsection, the continuous limit is taken by letting
 \beq
\sum_N=\sum_{nlm}\to \int_0^{\infty}dk\frac{dn}{dk}\sum_{l}\sum_m~.
 \eeq
Using the convention
 \beq
A_n^2=\frac{1}{2\pi}\frac{dk}{dn}~,\qquad \delta_{nm}=2\pi
  A_n^2\delta(k_n-k_m)~,
 \eeq
then
 \beq
\sum_{nlm}A_n^2\to\int_0^{\infty}\frac{dk}{2\pi}\sum_l\sum_m~,
 \eeq
and the normalization factor becomes $N_l(k)=2k$.
The expansion in terms of the eigenmode functions now reads
 \beq\label{modefunctions}
g(\vec{x})=\sum_{nlm}g_{nlm}u_{nlm}(\vec{x})=\sum_{nlm}g_{nlm}A_nN_lj_l(k_nr)Y_{lm}(\theta,\phi)~,
 \eeq
 where $u_{nlm}(\vec{x})$ is a solution to the equation of motion under the
  given boundary condition and the
set $k_n$ is determined by the boundary condition.

As an example, consider the Dirichlet boundary condition which yields
the constraint
 \beq
u_{nlm}(\vec{x}_B)= A_n N_l j_l(k_nr_B)Y_{lm}(\theta,\phi)=0~.
 \eeq
It is clear that the allowed $k_n$'s depend on $l$ through the
condition $j_l(k_nr_B)=0$. Thus, instead of Eq. (\ref{modefunctions})  we should write
 \beq
g(\vec{x})=\sum_{nlm}g_{nlm}A_{nl}N_lj_l(k_{nl}r)Y_{lm}(\theta,\phi)~,
 \eeq
 where $k_{nl}$ is now the physical momentum depending on
 both $n$ and $l$. Note that
in principle the amplitude $A$ can also depend on $l$.
From Eq. (\ref{knl}) we find
 \beq
A_{nl}^2=\frac{1}{2\pi}\frac{dk_{nl}}{dn}=\frac{1}{r_B}~.
 \eeq
Thus, very conveniently, for the Dirichlet boundary condition $A$
is actually independent of $l$ and $n$ in the large
argument limit.

\subsection{The discrete power spectrum}
The power
spectrum is defined as the power per log\,$k$ interval of the
variance $\sigma_g^2$, given by
 \bea
\nonumber \sigma_g^2(\vec{x})&\equiv &\left< g^2(\vec{x})\right>=\int d^3x~
g^2(\vec{x})\\
&=&\int d^3x~\sum_{nlm}\sum_{n'l'm'}\left< g^*_{nlm}g_{n'l'm'}\right>u^*_{nlm}(\vec{x})u_{n'l'm'}(\vec{x})=\sum_{nlm}\left<|g_{nlm}|^2\right>~.
 \eea
Using the relation
 \beq
g_{nlm}\equiv A_{nl} g_{lm}(k_{nl})=\frac{A_{nl}}{{\pi}N_l}k^2_{nl}i^l\int
 g(\vec{k}_{nl})Y_{lm}(\hat k_{nl})d\Omega_k~,
 \eeq
in the continuum limit one finds that
 \bea
\nonumber
& &\sigma_g^2(\vec{x})=\sum_{nlm}\left<|g_{nlm}|^2\right>\\
\nonumber
&\to&
 \int_0^{\infty}dk\sum_{lm}\frac{dn}{dk}A_{nl}^2\left<|g_{lm}(k)|^2\right>=\int_0^{\infty}\frac{dk}{2\pi}\sum_{lm}\left<|g_{lm}(k)|^2\right>\\
\nonumber
&=&\int_0^{\infty}\frac{dk}{2\pi}\sum_{lm}\frac{k^4}{\pi^2 N^2_l}\int\int\left<|g(\vec{k})|^2\right>Y_{lm}^*(\hat
 k')Y_{lm}(\hat k)d\Omega_kd \Omega_{k'}~.
 \eea
Using the orthonormality relation
 $
\sum_{lm}Y_{lm}(\theta,\phi)Y_{lm}^*(\theta',\phi')=\delta(\cos(\theta)-\cos(\theta'))\delta(\phi-\phi')
$
and using $N_l(k)=2k$, we obtain the correct continuum expressions
 \beq
\sigma_g^2(\vec{x})=\int_0^{\infty}\frac{dk}{(2\pi)^3}k^2
\int d\Omega_k\left<|g(\vec{k})|^2\right> = \int_0^{\infty}\frac{dk}{k}\mathcal{P}_g(k)
 \eeq
with
 \beq
\left<g^*(\vec{k})g(\vec{k}')\right>=(2\pi)^3\delta^3(\vec{k}-\vec{k}')\frac{2\pi^2}{k^3}\mathcal{P}_g(k)~.
 \eeq

\subsection{The Sachs-Wolfe effect}

The usual multipole expansion of the temperature anisotropies
reads
 \beq \frac{\delta T}{T_0}(\theta,\phi)=\sum
a_{lm}Y_{lm}(\theta,\phi)~,
 \eeq 
and the angular power spectrum is
given by
 \beq 
C_l\equiv \frac{1}{2l+1}\sum_m<|a_{lm}|^2>~.
 \eeq In
a matter dominated universe, the Sachs-Wolfe effect $\delta T/T =
-(1/5) \mathcal{R}$, implies
 \beq a_{lm}=\int d\Omega
Y_{lm}^*(\theta,\phi)\frac{\delta
T}{T_0}(\theta,\phi)=-\frac{1}{5}\int d\Omega_x Y_{lm}^*(\hat
x)\mathcal{R}(x_{dec}\hat x)~. \eeq Using \beq
\mathcal{R}(\vec{x}_{dec})=\sum_{nlm}\mathcal{R}_{nlm}A_{nl}N_{l}j_l(k_{nl}r)Y_{lm}(\theta,\phi)~,
 \eeq 
and the definition
 \beq 
\left< R^*_{n'l'm'}R_{nlm}\right> =
\frac{2\pi^2}{
k^3}\mathcal{P_R}(k)\delta_{nn'}\delta_{ll'}\delta_{mm'}~,
 \eeq 
we obtain by a straightforward standard calculation the discrete
version of the angular power spectrum as
 \beq C_l
=\frac{1}{25}\sum_{n}A_{nl}^2\frac{2\pi^2}{k_{nl}^3}\mathcal{P_R}(k_{nl})N_l^2j^2_l(k_{nl}r)~.
 \eeq 
Inserting $N_l^2=4k^2$ one obtains the usual continuum
result\footnote{Here the Fourier expansion normalization is chosen
to be $1/(2\pi)^3$; the $(2\pi)^{3/2}$ normalization would change
the power spectrum by a factor of $(2\pi)^{3}$.} $C_l
=\frac{4\pi}{25}\frac{1}{2\pi^2}\int\frac{dk}{k}j_l^2(kr)\mathcal{P_R}(k)~.
 $
with $A_{nl}^2=\frac{1}{2\pi}dk/dn$.

Note that the set of the allowed momenta $k$ in the discrete version of the Sachs-Wolfe
effect depends on $l$.

\section{A toy model with IR/UV cosmic duality}

In this section we outline the toy model of a cosmic CMB/dark energy
duality put forward in \cite{kems}. In a universe dominated by
dark energy in the asymptotic future we actually live inside a finite box,
the so-called "causal diamond" of the static de Sitter coordinates,
which is bounded by the past and future event horizons \cite{Banks:2000fe}. In
cosmic coordinates the finiteness could manifest itself as an
effective IR regulator
of the same order of magnitude as the  future event horizon, which in a
pure de Sitter space determines also the magnitude of the effective
cosmological constant.
If an IR/UV duality
is at work in the theory at some fundamental level, the IR
regulator might in some (complicated) way relate the dark energy and the
IR cutoff of the CMB perturbation modes. The model of \cite{kems},
that we explore here, might be considered as a simple toy model for
such a connection

Let us now assume that the size of the IR cutoff is related to the
the future event horizon $R_H=a\int_t^\infty {dt}/{a}$,
i.e. the effective size of the universe is the one that we can ever
hope to observe \cite{Banks:2000fe,Li:2004rb,Huang:2004ai,Huang:2004wt,kems}.
In other words, a local field theory should describe only the degrees of
freedom that can ever be observed by a given observer. In a universe
dominated by dark energy $R_H$ is of the order of the present
Hubble radius $H_0^{-1}$ but the actual value depends on the
equation of state of dark energy. We write
the IR cutoff $\tilde{L}$ as
\beq \label{ircut}
R_H\equiv c\tilde{L}~,
\eeq
where $c\sim {\cal O}(1)$ is a free parameter that will be related to dark energy.
The connection comes about by requiring that he
total energy in a region of spatial size $\tilde{L}$ should not
exceed the mass of a black hole of the same size, or
$\frac{4\pi}{3}\tilde{L}^3\rho_{\Lambda}\leq 4\pi\tilde{L}M_P^2$,
where $\Lambda$ is the UV cutoff. The largest IR cutoff $\tilde{L}$
is obtained by saturating the
inequality so that we write
 \beq \label{an}
\rho_{\Lambda}=3M_P^2 \tilde{L}^{-2}=3 c^2 M_P^2 R_H^{-2} ~.
 \eeq

We adopt a flat universe with $\Omega=1$. Then from the Friedmann equation
 and Eq. (\ref{an}) it follows that
 \beq
R_H =
a^{3/2}c\frac{1}{\sqrt{\Omega_m^0}H_0}\left(\frac{1-\Omega_{\Lambda}}{\Omega_{\Lambda}}\right)^{1/2}~,
 \eeq
which implies that today the value of the future event horizon would be
 \beq \label{RH}
R_H=\frac{c}{\sqrt{\Omega_{\Lambda}^{0}}}H_0^{-1}~,
 \eeq
 where the superscripts refer to the present values.

In flat space the multipole $l$ is given by
$l=k_l(\eta_0-\eta_*)$ where $\eta_0-\eta_*$ is the comoving
distance to the last-scattering surface and $k_l$ is the
corresponding comoving wave number. The comoving distance to last
scattering is given by
\beq
\label{eqeta} \eta_0-\eta_* =\int_0^{z_*}dz'\frac{1}{H(z')}
\eeq
and if the dominant energy components
are dark energy and matter, then
$H^2(z)=H_0^2\left[\Omega_{\Lambda}^0(1+z)^{(3+3w)}+(1-\Omega_{\Lambda}^0)(1+z)^3\right]~,$
where $w$ is given by the equation of state of dark energy
$p_{\Lambda}=w\rho_{\Lambda}$. Here $w$ is not a free parameter but
is related to the constant $c$ through \cite{Li:2004rb}
 \beq \label{eqofstate}
w = -\frac{1}{3}-\frac{2}{3c}\sqrt{\Omega_{\Lambda}}~.
 \eeq
This equation also holds at the present so that
 \beq \label{eospresent}
w_0 = -\frac{1}{3}-\frac{2}{3c}\sqrt{\Omega_{\Lambda}^0}~.
 \eeq

Thus, the distance to last scattering is a function of $w$ which
can be found by fixing the free parameter $c$. Since $\tilde L$ does not depend on $c$,
one finds with Dirichlet boundary condition \cite{kems}
$\tilde{L}=1.2\times H_0^{-1}$ or $k_{c}\equiv
\lambda_c/(2\pi)=1.2/\pi\times H_0^{-1}$, where $k_{c}$ defines the
smallest allowed wave number. But because the distance to last
scattering depends on $w$, the relative position of the cutoff in the
CMB spectrum depends on $w$.

Testing this model means fitting simultaneously the power spectrum and
$w$, using different boundary conditions. This yields the value of
$c$, which is the only free parameter here: the shape of the power
spectrum is fixed once $c$ (and the boundary condition) is fixed. Note
that because the space is effectively finite, there will be
oscillations at low $l$ in the power spectrum. These oscillations are
however not freely adjustable but depend on the equation of state of
dark energy, making the holographic toy model highly constrained and
predictive.

However, given that we do not know exactly how the boundary condition
should be imposed, we allow for more freedom in our fits to
data. Instead of fixing the infrared cut-off at $k_c\equiv
\lambda_c/(2\pi)=1.2/\pi\times H_0^{-1}$ we take it to be a free
parameter $k_{\rm cut}$. As it turns out the best fit is at $k_{\rm
cut} \sim 0.5-0.6 k_c$, and given our ignorance of how to impose the
boundary condition this is an approximation at the same level as
choosing different types of boundaries (Neumann, Dirichlet, etc.).
This leaves us with two parameters for any given model (in addition to
the choice of boundary condition): $c$ and $k_{\rm cut}$. In the
following section we present a likelihood analysis of the model using
CMB, LSS, and SNI-a data.

\section{Data analysis}

\subsection{The procedure}

In order to test the discrete models against data we employ the following procedure: First
a boundary condition (Dirichlet or Neumann in our case) is chosen. Next we calculate
$\chi^2$ for each set of $c$ and $k_{\rm cut}$, while marginalizing over all other cosmological parameters.
As our framework we choose the
minimum standard model with 6
parameters: $\Omega_m$, the matter density, $\Omega_b$, the baryon density, $H_0$, the Hubble
parameter, and $\tau$, the optical depth to reionization. The normalization of both
CMB and LSS spectra are taken to be free and unrelated parameters.
The priors we use are given in Table~\ref{tab:priors}.

\begin{table}[ht]
\label{tab:priors}
\caption{Priors on cosmological parameters used in
the likelihood analysis.}
\begin{center}
\begin{tabular}{@{}lll}
\hline
Parameter &Prior&Distribution\cr
\hline
$\Omega=\Omega_m+\Omega_X$&1&Fixed\\
$h$ & $0.72 \pm 0.08$&Gaussian \cite{freedman}\\
$\Omega_b h^2$ & 0.014--0.040&Top hat\\
$n_s$ & 0.6--1.4& Top hat\\
$\tau$ & 0--1 &Top hat\\
$Q$ & --- &Free\\
$b$ & --- &Free\\
\hline
\end{tabular}
\end{center}
\end{table}

Likelihoods are calculated from $\chi^2$ so that for 1 parameter estimates, 68\% confidence regions are determined by $\Delta \chi^2 = \chi^2 - \chi_0^2 = 1$, and 95\% region by $\Delta \chi^2 = 4$. $\chi_0^2$ is $\chi^2$ for the best fit model found.
In 2-dimensional plots the
68\% and 95\% regions are formally
defined by $\Delta \chi^2 = 2.30$ and 6.17
respectively. Note that this means that the 68\% and 95\% contours
are not necessarily equivalent to the same confidence level for single
parameter estimates.

\subsection{Description of the data sets}

\begin{itemize}
\item{Supernova luminosity distances}

We perform our likelihood analysis using the ``gold'' dataset compiled
and described in Riess et al \cite{Riess:2004} consisting of 157
SNe Ia using a modified version of the
SNOC package \cite{Goobar:2002c}.

\item{Large Scale Structure (LSS).}

At present there are two large galaxy surveys of comparable size, the
Sloan Digital Sky Survey (SDSS) \cite{Tegmark:2003uf,Tegmark:2003ud}
and the 2dFGRS (2~degree Field Galaxy Redshift Survey) \cite{2dFGRS}.
Once the SDSS is completed in 2005 it will be significantly larger and
more accurate than the 2dFGRS. In the present analysis we use data from SDSS, but the results would be almost identical had we used 2dF data instead. In the data analysis we use only data points on scales larger than $k = 0.15 h$/Mpc in order to avoid problems with non-linearity.

\item{Cosmic Microwave Background.}

The CMB temperature fluctuations are conveniently described in terms of
the spherical harmonics power spectrum $C_l^{TT} \equiv \langle
|a_{lm}|^2 \rangle$, where $\frac{\Delta T}{T} (\theta,\phi) =
\sum_{lm} a_{lm}Y_{lm}(\theta,\phi)$.  Since Thomson scattering
polarizes light, there are also power spectra coming from the
polarization. The polarization can be divided into a curl-free $(E)$
and a curl $(B)$ component, yielding four independent power spectra:
$C_l^{TT}$, $C_l^{EE}$, $C_l^{BB}$, and the $T$-$E$ cross-correlation
$C_l^{TE}$.

The WMAP experiment has reported data only on $C_l^{TT}$ and $C_l^{TE}$
as described in
Refs.~\cite{Bennett:2003bz,Spergel:2003cb,%
Verde:2003ey,Kogut:2003et,Hinshaw:2003ex}.  We have performed our
likelihood analysis using the prescription given by the WMAP
collaboration~\cite{Spergel:2003cb,%
Verde:2003ey,Kogut:2003et,Hinshaw:2003ex} which includes the
correlation between different $C_l$'s. Foreground contamination has
already been subtracted from their published data.
\end{itemize}

\subsection{Results}

\begin{figure}
\hspace*{1.5cm}\includegraphics[width=100mm]{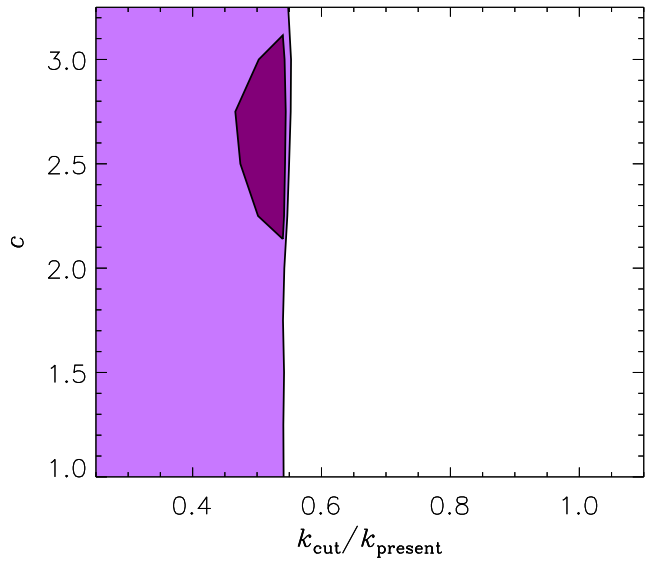}
\caption{68\% (dark) and 95\% (light) likelihood contours for WMAP and SDSS data using the Dirichlet boundary condition.}
\label{fig1}
\end{figure}

First, we perform the likelihood analysis separately for the CMB+LSS
data and the SNIa data.  In Fig.~\ref{fig1} we show the likelihood in
$c,k_{\rm cut}$ space for the Dirichlet boundary condition. The best
fit model \footnote{Note that in the holographic toy model, the
number of degrees of freedom is the same as in the $\Lambda$CDM
model since $k_{cut}$ and $w$ are both determined by $c$.} has
$\chi^2 = 1444.8$ as opposed to the best fit $\Lambda$CDM model which
has $\chi^2 = 1447.5$.

\begin{figure}
\hspace*{1.5cm}\includegraphics[width=100mm]{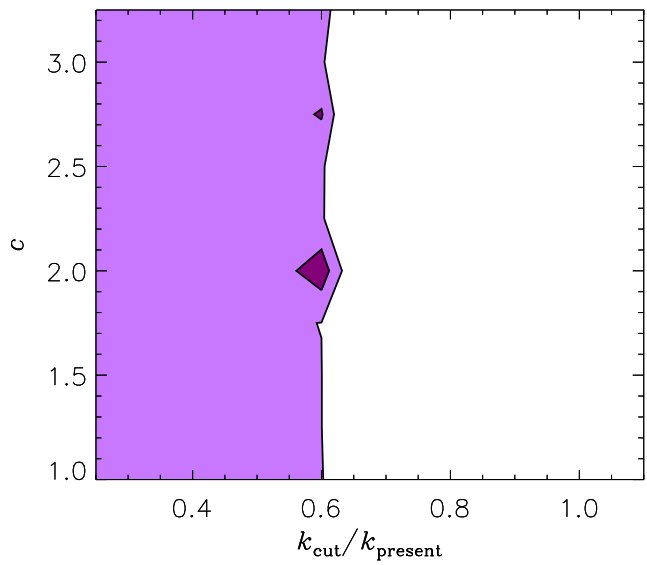}
\caption{68\% (dark) and 95\% (light) likelihood contours for WMAP and SDSS data using the Neumann boundary condition.}
\label{fig2}
\end{figure}

In Fig.~\ref{fig2} we show the same analysis, but for the Neumann boundary condition. Here,
the best fit is $\chi^2 = 1441.4$, somewhat better than for the previous case.

\begin{figure}
\hspace*{1.5cm}\includegraphics[width=100mm]{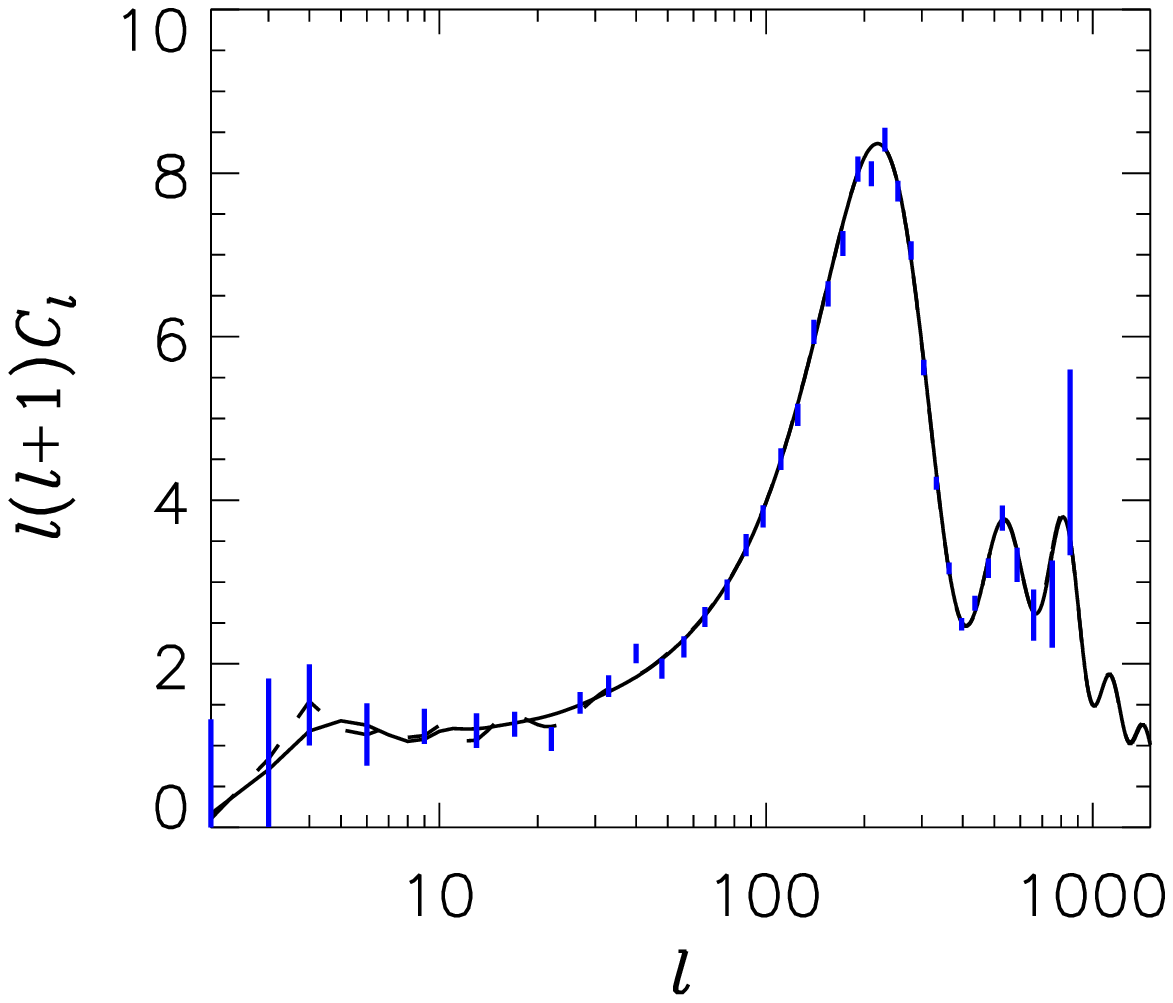}
\caption{Temperature power spectra for the two best fit models, the full line is with Dirichlet
boundary conditions and the dashed with Neumann boundary conditions. The data shown is the binned WMAP data.}
\label{fig3}
\end{figure}

In Fig.~\ref{fig3} we show CMB temperature power spectra for the two
best fit models \cite{CMBFAST}.
Interestingly, the curve with Neumann boundary conditions is able to fit the small $l$
spectrum oscillations almost perfectly, including the $l \sim 20$ feature. However,
the $l \sim 40$ feature is not reproduced. In both cases the power spectrum
suppression at small $l$ is reproduced nicely because of the large scale cut-off in the spectrum.

\begin{figure}
\hspace*{1.5cm}\includegraphics[width=100mm]{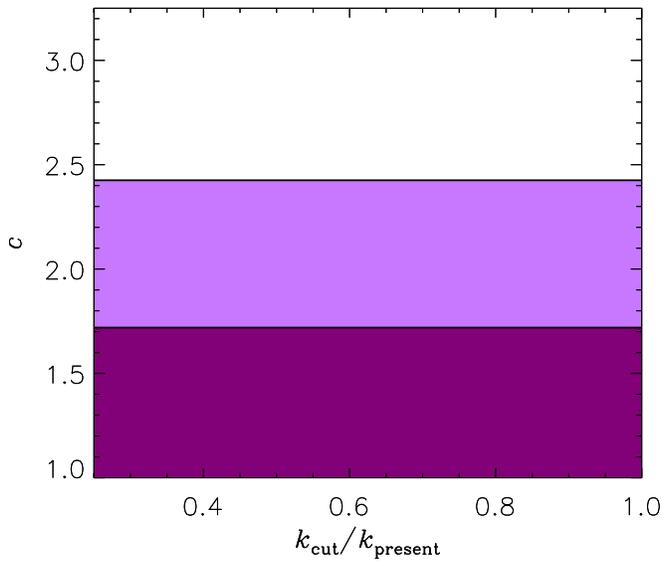}
\caption{68\% (dark) and 95\% (light) likelihood contours for SNIa data.}
\label{fig4}
\end{figure}

Fig.~\ref{fig4} show the same analysis performed for the SNIa data. In this case there is no
dependency on the cut-off scale, only on $c$, because that parameter modifies the dark energy
behaviour. The best fit is at $\chi^2 = 177.1$ which is almost identical to that of the
best fit $\Lambda$CDM model.

\begin{figure}
\hspace*{1.5cm}\includegraphics[width=100mm]{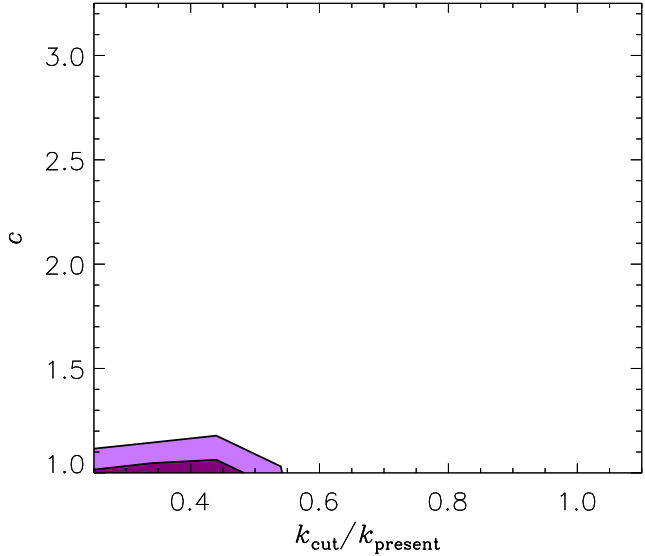}
\caption{68\% (dark) and 95\% (light) likelihood contours for WMAP, SDSS and SNIa data.}
\label{fig5}
\end{figure}

However, as can be seen, the best fits of CMB+LSS and SNIA
are incompatible. This is illustrated in Fig.~\ref{fig5} which shows a combined
CMB+LSS+SNIa fit for the Dirichlet boundary condition. Here, the best fit is
at $\chi^2 = 1636.8$, as opposed to the best fit $\Lambda$CDM model which has $\chi^2 = 1626.4$.
The fundamental reason for the discrepancy is just the well-known degeneracy between the matter density
and the dark energy equation of state, $w$. If only CMB and LSS data is considered,
then having $w > -1$ generally requires a higher matter density, whereas for SNIa
data the reverse is true. This means that the combination in general rules out
any model with a $w$ which is too high. For a constant $w$ the present bound is
$w \leq -0.79$ at 95\% C.L. \cite{steenmortsell}.

\section{Discussion}

It is intriguing that with the holographic toy model of Sect. 4
the power spectrum and in particular the low $l$ features are automatically fitted almost perfectly, as
can be gathered from Fig. 3. The CMB and LSS data, on its own, would indeed
be consistent with an IR/UV cosmic duality with an IR cutoff.
However, from the analysis that combined CMB and LSS with the supernova data
the conclusion clearly seems to be that in its present form the toy
model of cosmic duality is strongly disfavored.

In fact, it would have been a great surprise if such a simple toy
model would have worked perfectly as the actual holographic constraint
is expected to be rather complicated.
It is even conceivable that it
manifests itself in different ways at different
length scales. From a holographic point of view, supernova luminosity
effects represent
ordinary local physics and therefore have a different status
from dark energy measured by CMB, which is a more global concept.
It is also possible that the
apparent decrease in the distant supernova luminosities is not due to
accelerated expansion but to some other physical process.  An example
would be the recently suggested axion-photon mixing \cite{kaloper}.

That said, one should note that if we use, for the cutoff on the large
scale CMB anisotropies, the future
event horizon at last scattering \cite{kems}, then with $c=2$ and
$\Omega_{\Lambda}^0=0.7$ corresponding to $w_0=-0.61$, the toy
model with CMB/dark energy duality implies
$k_{cut}/k_{present}=0.6$. Curiously that is in the maximum likelihood
region with Neumann boundary conditions (see Fig.~2).

The present holographic model presents one
possibility of how new physics might effect present day
cosmology; some others have been discussed in light of observational data in \cite{Hannestad:2004ts}. Although
no firm conclusions can be drawn at this stage, it is nevertheless
encouraging that data is good enough for testing these ideas.
Signals for a discrete power spectrum and a holographic connection
between the ultraviolet and the infrared remain interesting
possibilities linked to fundamental physics that are
worth searching for also in the forthcoming cosmological data.

\section*{Acknowledgments}

SH and MSS wish to thank the CERN theory division for hospitality and
support. KE is partly supported by the Academy of Finland grants 75065
and 205800. The work of MSS was supported in part by the DOE Grant
DE-FG03-91ER40674. In addition MSS would like to acknowledge useful discussions
with Nemanja Kaloper and Archil Kobakhidze.

\section*{References} 


\begin{thebibliography}{99}

\bibitem{Bennett:2003bz}
C.~L.~Bennett {\it et ll.},
Astrophys.\ J.\ Suppl.\  {\bf 148}, 1 (2003)
[arXiv:astro-ph/0302207];

\bibitem{Spergel:2003cb}
D.~N.~Spergel {\it et al.}  [WMAP Collaboration],
Astrophys.\ J.\ Suppl.\  {\bf 148}, 175 (2003)
[arXiv:astro-ph/0302209].



\bibitem{topolimits}
N.~J.~Cornish, D.~N.~Spergel, G.~D.~Starkman and E.~Komatsu,
Phys.\ Rev.\ Lett.\  {\bf 92}, 201302 (2004)
[arXiv:astro-ph/0310233];
N.~G.~Phillips and A.~Kogut,
arXiv:astro-ph/0404400.

\bibitem{holo}
G.~'t Hooft,
gr-qc/9310026.
L.~Susskind,
J.\ Math.\ Phys.\  {\bf 36}, 6377 (1995) [hep-th/9409089];
R.~Bousso,
Rev.\ Mod.\ Phys.\  {\bf 74}, 825 (2002) [hep-th/0203101].



\bibitem{bekenstein}
J.~D.~Bekenstein,
Phys.\ Rev.\ D {\bf 7} (1973) 2333.

\bibitem{Banks:2001px}
A.~Strominger,
JHEP {\bf0111}, 049 (2001)
[hep-th/0110087];
V.~Balasubramanian, J.~de Boer and D.~Minic,
Phys.\ Rev.\ D {\bf 65}, 123508 (2002)
[hep-th/0110108];
T.~Banks and W.~Fischler,
hep-th/0111142;
C.~J.~Hogan,
Phys.\ Rev.\ D {\bf 66}, 023521 (2002)
[astro-ph/0201020];
F.~Larsen, J.~P.~van der Schaar and R.~G.~Leigh,
JHEP {\bf 0204}, 047 (2002)
[hep-th/0202127];
A.~Albrecht, N.~Kaloper and Y.~S.~Song,
hep-th/0211221;
U.~H.~Danielsson,
JCAP {\bf 0303}, 002 (2003)
[hep-th/0301182];
E.~Keski-Vakkuri and M.~S.~Sloth,
JCAP {\bf 0308}, 001 (2003)
[hep-th/0306070];
J.~P.~van der Schaar,
JHEP {\bf 0401}, 070 (2004)
[hep-th/0307271];
T.~Banks and W.~Fischler,
hep-th/0310288;
C.~J.~Hogan,
astro-ph/0310532;
F.~Larsen and R.~McNees,
hep-th/0402050.


\bibitem{Banks:2000fe}
T.~Banks,
hep-th/0007146;
R.~Bousso,
JHEP {\bf 0011}, 038 (2000)
[hep-th/0010252];
T.~Banks and W.~Fischler,
hep-th/0102077;
L.~Dyson, M.~Kleban and L.~Susskind,
JHEP {\bf 0210}, 011 (2002)
[hep-th/0208013];
M.~K.~Parikh, I.~Savonije and E.~Verlinde,
Phys.\ Rev.\ D {\bf 67}, 064005 (2003)
[hep-th/0209120];
N.~Kaloper, M.~Kleban, A.~Lawrence, S.~Shenker and L.~Susskind,
JHEP {\bf 0211}, 037 (2002)
[hep-th/0209231];
T.~Banks, W.~Fischler and S.~Paban,
JHEP {\bf 0212}, 062 (2002)
[hep-th/0210160];
U.~H.~Danielsson, D.~Domert and M.~Olsson,
Phys.\ Rev.\ D {\bf 68}, 083508 (2003)
[hep-th/0210198];
A.~Albrecht and L.~Sorbo,
hep-th/0405270;
B.~Freivogel and L.~Susskind,
arXiv:hep-th/0408133.

\bibitem{Hsu:2004ri}
S.~D.~H.~Hsu,
hep-th/0403052.

\bibitem{Li:2004rb}
M.~Li,
hep-th/0403127.

\bibitem{Huang:2004ai}
Q.~G.~Huang and M.~Li,
astro-ph/0404229.

\bibitem{Huang:2004wt}
Q.~G.~Huang and Y.~G.~Gong,
astro-ph/0403590.

\bibitem{kems}
K.~Enqvist and M.~S.~Sloth,
Phys.\ Rev.\ Lett.\  {\bf 93}, 221302 (2004) [hep-th/0406019].

\bibitem{Cohen:1998zx}
A.~G.~Cohen, D.~B.~Kaplan and A.~E.~Nelson,
Phys.\ Rev.\ Lett.\  {\bf 82}, 4971 (1999) [hep-th/9803132].


\bibitem{Thomas:pq}
S.~Thomas,
Phys.\ Rev.\ Lett.\  {\bf 89}, 081301 (2002).

\bibitem{'tHooft:1984re}
G.~'t Hooft,
Nucl.\ Phys.\ B {\bf 256}, 727 (1985).

\bibitem{Susskind:1993if}
L.~Susskind, L.~Thorlacius and J.~Uglum,
Phys.\ Rev.\ D {\bf 48}, 3743 (1993)
[arXiv:hep-th/9306069].

\bibitem{Tocchini-Valentini:2004ht}
D.~Tocchini-Valentini, M.~Douspis and J.~Silk,
arXiv:astro-ph/0402583.


\bibitem{transplanck}
J.~Martin and C.~Ringeval,
Phys.\ Rev.\ D {\bf 69}, 083515 (2004)
[arXiv:astro-ph/0310382];
J.~Martin and C.~Ringeval,
Phys.\ Rev.\ D {\bf 69}, 127303 (2004)
[arXiv:astro-ph/0402609].

\bibitem{Hannestad:2004ts}
S.~Hannestad and L.~Mersini-Houghton,
hep-ph/0405218 (to appear in Phys. Rev. D).


\bibitem{huntsarkar}
P.~Hunt and S.~Sarkar,
arXiv:astro-ph/0408138.


\bibitem{Hannestad:2003zs}
S.~Hannestad,
JCAP {\bf 0404}, 002 (2004)
[astro-ph/0311491].

\bibitem{steenmortsell}
S.~Hannestad and E.~Mortsell,
JCAP {\bf 0409}, 001 (2004) [astro-ph/0407259].


\bibitem{freedman}
W.~L.~Freedman {\it et al.},
Astrophys.\ J.\  {\bf 553}, 47 (2001)
[arXiv:astro-ph/0012376].




\bibitem{2dFGRS} M.~Colless {\it et al.},
astro-ph/0306581.

\bibitem{Tegmark:2003ud}
M.~Tegmark {\it et al.}  [SDSS Collaboration],
astro-ph/0310723.

\bibitem{Tegmark:2003uf}
M.~Tegmark {\it et al.}  [SDSS Collaboration],
astro-ph/0310725.

\bibitem{Verde:2003ey}
L.~Verde {\it et al.},
Astrophys.\ J.\ Suppl.\ {\bf 148} (2003) 195
[astro-ph/0302218].


\bibitem{Kogut:2003et}
A.~Kogut {\it et al.},
Astrophys.\ J.\ Suppl.\ {\bf 148} (2003) 161
[astro-ph/0302213].

\bibitem{Hinshaw:2003ex}
G.~Hinshaw {\it et al.},
Astrophys.\ J.\ Suppl.\ {\bf 148} (2003) 135
[astro-ph/0302217].

\bibitem{CMBFAST}
U.~Seljak and M.~Zaldarriaga,
Astrophys.\ J.\  {\bf 469} (1996) 437
[astro-ph/9603033].
See also the CMBFAST website at
{\tt http://www.cmbfast.org}

\bibitem{Riess:2004}
A.~G.~Riess {\it et al.}  [Supernova Search Team Collaboration],
Astrophys.\ J.\  {\bf 607}, 665 (2004)
[arXiv:astro-ph/0402512].


\bibitem{Goobar:2002c}
Goobar, A., M\"ortsell, E., Amanullah, R.,
Goliath, M., Bergstr\"om, L. and Dahl\'en, T., 2002,
Astron.\ Astrophys., 392, 757
[astro-ph/0206409]. Code available at {\tt
http://www.physto.se/$\sim$ariel/snoc/}

\bibitem{kaloper}
C.~Csaki, N.~Kaloper and J.~Terning,
Phys.\ Rev.\ Lett.\  {\bf 88}, 161302 (2002)
[hep-ph/0111311];
C.~Deffayet, D.~Harari, J.~P.~Uzan and M.~Zaldarriaga,
Phys.\ Rev.\ D {\bf 66}, 043517 (2002)
[arXiv:hep-ph/0112118];
C.~Csaki, N.~Kaloper and J.~Terning,
Phys.\ Lett.\ B {\bf 535}, 33 (2002)
[arXiv:hep-ph/0112212];
E.~Mortsell, L.~Bergstrom and A.~Goobar,
Phys.\ Rev.\ D {\bf 66}, 047702 (2002)
[arXiv:astro-ph/0202153];
M.~Christensson and M.~Fairbairn,
Phys.\ Lett.\ B {\bf 565}, 10 (2003)
[arXiv:astro-ph/0207525];
E.~Mortsell and A.~Goobar,
JCAP {\bf 0304}, 003 (2003)
[arXiv:astro-ph/0303081].





\end{thebibliography}
\end{document}